\documentstyle[prb,aps, twocolumn]{revtex}
\newcommand{\rL}{\rho_{\scriptscriptstyle L}}
\newcommand{\vel}{{\bf v}_{\scriptscriptstyle L}}
\newcommand{\scs}{\scriptscriptstyle}
\newcommand{\veli}{v_{{\scriptscriptstyle L}i}}
\newcommand{\velc}{(\vel\cdot\nabla)}
\newcommand{\tsi}{\tilde{\sigma}}
\newcommand{\be}{\begin{equation}}
\newcommand{\ee}{\end{equation}}
\newcommand{\onu}{\overline{\nu}}
\begin{document}
\draft
\title{Hydrodynamics of crystals with interstitials}
\author{G.L Buchbinder\cite{byline}}
\address{Department of Physics, Omsk State University, Peace Avenue 55-a 644077, Omsk, Russia}
\date{\today}
\maketitle
\begin{abstract}
The hydrodynamic equations for a crystals with interstitials, taking
into account the dissipative processes of the viscosity, heat conduction
and the interstitial diffusion are derived. To achieve that
we use the phenomenological approach has originally been applied in the theory of the
superfluids for the derivation of equations of the two-fluid hydrodynamics.
On the basis of obtained equations the problem of the propagation of plane waves
in the crystal with low interstitial concentration has been considered. For the case
when effects of the viscosity and the heat conduction are absent and the diffusion
mobility of interstitials is small the absorption coefficient of longitudinal  sound wave has been
calculated.
\end{abstract}
A great number of various in its nature processes occuring in the bulk and on the
surface of the crystalline solids can be considered within the scope of the hydrodynamic
description
\cite{M72,Fl76,A81,L85,AP92,G92,AM95,S96,W96,S98,T99}.
The presence of the slowly varying spatial and
temporal disturbances in the system allows us to describe the nonequilibrium behavior
in terms of a few slow, hydrodynamic variables. One of the most important hydrodynamic
processes is the diffusion of interstitials in the crystalline lattice. Within the scope
of the hydrodynamic approximation their dynamics is commonly determined by the
well known diffusion equation rested on the classical Fick's laws. This equation
is closed with respect to the interstitial concentration and the presence of
the lattice is only manifested at the calculation of the diffusion coefficient
for the specific microscopic model. Such an approach, justified in the case
of the low impurity concentrations and weak inhomogeneity in the system  proves
to be unfit at the finite concentration and  the large gradients. In the case
the interaction  between the impurity system and the lattice is also manifested
at the macroscopic level through the deformation of the crystal. Therefore
at the finite interstitial concentration the dynamics of interstitials is to be
considered in common with motion of the lattice. As a result the hydrodynamic
equations describing the corresponding process are to contain both the interstitial and
lattice variables.

In this paper we shall obtain hydrodynamic equations for the crystal containing
interstitials. With that end  in view we use the method has originally been applied
in the theory of the superfluids
\cite{L87,Kh71}
for the derivation of the equations of
the two-fluid hydrodynamics and then  employed for supersolids \cite{A69}.
It gives a possibility to obtain equations describing the dynamics of interstitials
relative to the moving lattice.

On the basis of the found equations we shall consider the propagation of plane waves in the crystal and
in the simplest approximation calculate the sound velocities and absorption
coefficient caused by a diffusion of interstitials.

We assume that the concentration of vacancies is negligible and the
nonequilibrium state of the crystal is characterized by the
mass densities of impurity particles $\rho_p$ and lattice atoms $\rho_{\scriptscriptstyle L}$
and denote by ${\bf j}$ the current density of the medium (lattice plus interstitials).
\begin{equation}
{\bf j} =  \rho_{\scriptscriptstyle L}{\bf v}_{\scriptscriptstyle L} + {\rho_p}{\bf v}_p\, ,
\label{1}
\end{equation}
where ${\bf v}_{\scriptscriptstyle L}$ and  ${\bf v}_p$ are the velocity fields of the
lattice and the impurity system, respectively.

The complete set of the hydrodynamic equations has to contain the local
conservation laws of mass, momentum and equation for the entropy, following from
the second law of the thermodynamics.  We write down the corresponding
balance equations as
\begin{eqnarray}
&&\frac{\partial\rho}{\partial t} + \mbox{div}\, {\bf j}  =  0\, ,\label{2}\\
&&\frac{\partial j_i}{\partial t} + \frac{\partial\Pi _{ik}}{\partial x_k}  =  0\, ,\label{3}\\
&&\frac{\partial S}{\partial t} + \mbox{div}\, (S{\bf v} + \frac{{\bf q}}{T} - \frac{\nu}{\rho T}{\bf J}) = \frac{R}{T}\, ,
\label{4}
\end{eqnarray}
here $\rho = \rho_{\scriptscriptstyle L} + {\rho_p}$ is the total mass density of the medium,
${\bf v} = \rho^{-1}{\bf j}$ is the mass velocity, $S$ is the entropy density of the medium,
$\Pi_{ik}$ is the tensor momentum current density, ${\bf q}$ is the
heat current density, ${\bf J}$ is the diffusion current density of interstitials
defined as
\[{\bf J} = \rho_p({\bf v}_p - {\bf v})\, , \]
$\nu$ is the chemical potential of interstitials per unite volume, $T$ is
absolute temperature  and $R\hspace{0.2cm} (R > 0)$ is the dissipative function of the medium;
summation over the repeated indexes is implied.

In addition, the system of equations (\ref{2}) - (\ref{4}) has to be supplemented by
the continuity equation for the interstitial density and equation of motion for the
lattice
\begin{eqnarray}
&&\frac{\partial(c\rho)}{\partial t} + \mbox{div}\, c\rho{\bf v}_p  =  0\, ,\label{5} \\
&&\frac{\partial(\rho_{\scriptscriptstyle L}v_{{\scriptscriptstyle L}{i}})}{\partial t} +
\frac{\partial\Pi _{{\scriptscriptstyle L}{ik}}}{\partial x_k} =  f_i\, ,\label{6}
\end{eqnarray}
where $c =   \rho_p/\rho$ is the interstitial concentration, $\bf f$ is the mass force with which
the impurity system acts on the lattice; $\Pi_{{\scs L}ik}$ is the tensor momentum current
density of the lattice which we take in the form
\begin{equation}
\Pi_{{\scs L}ik} = \rL\veli v_{{\scs L}k} - \sigma_{ik} - \sigma^{\prime}_{ik}\, ,
\label{7}
\end{equation}
where  $\sigma_{ik}$ is the symmetric, elastic stress tensor of the lattice
\begin{eqnarray}
\sigma_{ik}&=&-p{\scs L}\delta_{ik} + \tilde{\sigma}_{ik}\, , \\
p_{\scs L} &=& -\frac{1}{3}\sigma_{ii},\hspace{0.5cm}\mbox{Tr}\,\tilde{\sigma} = \tilde{\sigma}_{ii} = 0\, ,   \nonumber
\label{8}
\end{eqnarray}
and $\sigma^{\prime}_{k}$ describes the effect of the viscosity. Both $\sigma_{ik}$ and
$\sigma^{\prime}_{ik}$
are supposed to be known. Here and further the tilde is used to denote the traceless part of a tensor.

Now let us introduce the vector  $\bf u$ defining displacement of lattice sites
and connected with velocity field $\vel$ by relationship
\begin{equation}
\vel = \frac{d{\bf u}}{d\, t} = \frac{\partial{\bf u}}{\partial t} + (\vel\cdot\nabla){\bf u}\, .
\label{9}
\end{equation}

The variables $\rho$, $S$, $T$, $c$, $\bf j$, $\bf u$ completely define the nonequilibrium state of the system
and satisfy equations  (\ref{2}) - (\ref{6}).  Similarly to Ref. \cite{L87} we shall
find the remaining unknown values $\Pi_{ik}$, $R$, ${\bf q}$, $\bf J$, $\bf f$
so that the conservation
energy law
\begin{equation}
\frac{\partial E}{\partial t} + \mbox{div}\,{\bf Q}  =  0\, ,
\label{10}
\end{equation}
with $E$ and ${\bf Q}$ being energy density and energy current density, respectively, would follow
from Eqs.(\ref{2}) - (\ref{6}). In the Eq.(\ref {10}) the current $\bf Q$ is originally
unknown as well.

To determine the form of the unknown values indicated above we shall pass
to the new frame moving
with velocity $\vel$, in which the velocity of the lattice of the given element of the medium is
equal zero. The energy $E$ and the momentum density $\bf j$  are related by the Gallilean
transformation to its values $E_0$ and ${\bf j}_0$ in the frame where  the lattice
rests by relationships
\begin{equation}
E = \frac{\rho{v}_{\scs L}^2}{2} + {\bf j}_0{\vel} + E_0\, ,
\label{11}
\end{equation}
\begin{equation}
{\bf j} = \rho{\vel} + {\bf j}_0\,  ; \hspace{0.5cm}{\bf j}_0 = {\rho}_p({\bf v}_p - {\vel})\, .
\label{12}
\end{equation}

Let us write down the differential of $E_0$, considered as a function of $S$, $\rho$,
$c$, ${\bf j}_0$ and the infinitesimal strain tensor $u_{ij}$, in the form
\begin{equation}
dE_0 = TdS + \mu d\rho + \nu d c + \tilde{\sigma}_{ij}d\tilde{u}_{ij} + {\bf w}d{\bf j}_0 \, ,
\label{13}
\end{equation}
here $\mu$ is the chemical potential of the medium and ${\bf w} = {\bf v}_p - \vel$ is
the relative velocity. In the Eq. (\ref{13}) it
has been taken into account that the variation $du_{ii}$ of sum of diagonal components of the
strain tensor $u_{ij}$ is determined by the variation of the density $d\rho$.

Differentiating the Eq.(\ref{11}) with respect to time and using  Eqs.
(\ref{2}) - (\ref{13}),
one obtains
\begin{eqnarray}
\frac{\partial E}{\partial t}& = & ({\bf v}_p\vel -\frac{{v}_{\scs L}^2}{2} -
\mu + \frac{c\nu}{\rho})\,\mbox{div}\, {\bf j} - {\bf w}{\bf f}   \nonumber  \\
& &+\, \rL w_i\velc \veli - w_i\frac{\partial }{\partial x_k}(\sigma_{ik} + \sigma^{\prime}_{ik}) \nonumber \\
& & -\, v_{pi}\frac{\partial {\Pi}_{ik}}{\partial x_k} + \tsi_{ij}\frac{\partial u_{ij}}{\partial t} + \frac{\nu}{\rho}\mbox{div}\, c\rho{\bf v}_p  \nonumber  \\
& & -\, T\mbox{div}(S{\bf v} + \frac {{\bf q}}{T} - \frac{\nu}{\rho T}{\bf J})  + R \nonumber
\end{eqnarray}
The last equality after long and tedious, though simple, transformations is led to
the form
\begin{eqnarray}
& &\frac{\partial E}{\partial t} + \mbox{div} \Bigl\{ \big (\frac{v_{\scs L}^2}{2} + \mu + \frac{TS}{\rho}\big )\, {\bf j} + \rho_p{\bf v}_p({\bf v}_p\cdot{\bf w})  \nonumber \\
& & \hspace{1.5cm}+\, {\bf q} - \vel\cdot(\tsi + {\bf \sigma}^{\prime}) + {\bf v}_p\cdot{\bf\pi}  \Bigr \} \nonumber\\
& & \hspace{0.5cm}= \Big[R- \sigma^{\prime}_{ik}\frac{\partial \veli}{\partial x_k} + \pi_{ik}\frac{\partial v_{pi}}{\partial x_k}
+ \frac{{\bf q}\nabla T}{T}  \nonumber \\
& & \hspace{3.8cm}+ T{\bf J}\nabla\Big(\frac{\nu}{\rho T} \Big)\Big]  \nonumber\\
& & \hspace{1cm}\,- {\bf w}\Big[{\bf f} - \nabla p_{\scriptscriptstyle L} + \rL(\nabla\mu +
\frac{S\nabla T}{\rho})  \nonumber\\
& & \hspace{2.5cm} - \rL\frac{\nu}{\rho}\nabla\, c + {\tilde u}_{ij}\nabla\tsi_{ij} \Big] \, ,
\label{14}
\end{eqnarray}
where $\pi_{ik}$ is defined from the equality
\begin{eqnarray}
\Pi_{ik}& = &\rL\veli v_{{\scs L}k} +\rho_p v_{pi} v_{pk} + p\delta_{ik}   \nonumber\\
	&   & \phantom{\rL\veli v_{{\scs L}k} + \rho_p}-\tsi_{ik} -  \sigma^{\prime}_{ik} + \pi_{ik}
\label{15}
\end{eqnarray}
and the notation has been introduced
\begin{equation}
p = -E_0 + TS +\mu\rho + {\bf w}\cdot{\bf j}_0 + \tsi_{ij}{\tilde u}_{ij}
\label{16}
\end{equation}
In the derivatiopn of Eq.(\ref{14}) we have neglected the term $\tsi_{ij}{\tilde u}_{ik}\partial v_{{\scs L}k}/\partial x_j $ that
is small in framework of the linear elasticity theory. In addition, for the terms
of the second order of infinitesimal in strains the approximation has been used

\[({\bf v}_p - \vel)\cdot\nabla(\tsi_{ik}{\tilde u}_{ik}) \simeq {\bf v}_p\cdot\nabla(\tsi_{ik}{\tilde u}_{ik})\, ,\]
because the interstitial velocity is appreciably more than the lattice one.

The Eq.(\ref 15) defines the momentum current density of the medium. The comparison of the Eq.(\ref{14}) with the energy conservation law (\ref{10})
leads to the definition of: \newline the energy current ${\bf Q}$
\begin{eqnarray}
{\bf Q} & = & \Bigl\{ \big (\frac{v_{\scs L}^2}{2} + \mu + \frac{TS}{\rho}\big )\, {\bf j} + \rho_p{\bf v}_p({\bf v}_p\cdot{\bf w})\nonumber\\
& & +\, {\bf q} - \vel\cdot(\tsi + {\bf \sigma}^{\prime}) + {\bf v}_p\cdot{\bf\pi}  \Bigr \} \, ;
\label{17}
\end{eqnarray}
the dissipative function $R$
\begin{eqnarray}
 R = \sigma^{\prime}_{ik}\frac{\partial \veli}{\partial x_k} - \pi_{ik}\frac{\partial v_{pi}}{\partial x_k}
& - & \frac{{\bf q}\nabla T}{T}  \nonumber\\
& - &\, T{\bf J}\nabla\Big(\frac{\nu}{\rho T}\Big)\, ;\label{18}
\end{eqnarray}
and the force ${\bf f}$
\begin{eqnarray}
{\bf f} =  \nabla p_{\scs L} - \rL ({\nabla\mu} + \frac{S\nabla T}{\rho}) & + & {\rL}{\frac{\nu}{\rho}}\nabla\, c  \nonumber\\
& - & {\tilde u}_{ik}\nabla \tsi_{ik} \, .  \label{19}
\end{eqnarray}
In the framework of the linear theory the positive definiteness of $R$ leads to the
linear relationships relating the dissipative currents to the thermodynamic forces.
Taking into account Onsager's reciprocity relations for the transport coefficients
and time-reversal property of dissipative effects \cite{M72}, we can write this
relationships, at given $\sigma^{\prime}_{ik}$, in the form
\begin{eqnarray}
\pi_{ik} & = & -  \eta_{ikjm}\frac{\partial v_{pj}}{\partial x_m}\, , \nonumber \\
 q_i & = & - \frac{\kappa_{ik}}{T}\frac{\partial T}{\partial x_k} - \alpha_{ik}T\frac{\partial }{\partial x_k}\Big( \frac{\nu}{\rho T}\Big)\, , \label{20}\\
 J_i & = & -  \frac{\alpha_{ik}}{T}\frac{\partial T}{\partial x_k} -  \beta_{ik}T\frac{\partial}{\partial x_k}\Big( \frac{\nu}{\rho T}\Big)\, ,                \nonumber
\end{eqnarray}
where the fourth-rank tensor ${\bf\eta}$ is related to the impurity viscosity, the second-rank tensor
$\kappa$ has meaning of the pure heat conductivity, ${\bf \beta}$ is the second-rank
tensor related to the interstitial diffusion  and ${\bf \alpha}$ is related
to the cross effect of thermal diffusion of interstitials.

It is seen from the Eq.(\ref{15}) that $p$, given by the Eq.(\ref{16}),
can be interpreted as "pressure" in the medium. The relation similar to the Gibbs-Duhem
relation for fluid system follows from Eq.(\ref{16}).
\be
\rho d\mu = dp + \nu d\, c - SdT -{\bf j}_0d{\bf w} - {\tilde u}_{ij}d\tsi_{ij}\, .\label{21}
\ee
From (\ref{21}) one obtains for gradients
\[\rho\nabla\mu + S\nabla T = \nabla p + \nu\nabla c - j_{0i}\nabla w_i  -  {\tilde u}_{ij}\nabla\tsi_{ij} \, . \]
Introducing this into the Eq.(\ref{19}), we obtain the expression for the mass
force as
\begin{equation}
{\bf f} = \nabla p_{\scs L} - (1-c)\nabla p + c\rL \nabla \frac{w^2}{2} - c{\tilde u}_{ij}\nabla\tsi_{ij}\, .
\label{22}
\end{equation}
This expression shows explicitly that in the limit $c \rightarrow 0$ the mass
force vanishes provided that $p\rightarrow p_{\scs L}$.
From this condition it follows that it must be
\[p_{\scs L} = -E_0 + TS +\mu\rL + \tsi_{ij}{\tilde u}_{ij}, \hspace{0.5cm(c = 0)}. \]
The last relationship should be considered as the definition of the "pressure" $p_{\scs L}$ . Let us note
that the similar  expression has been used in Ref.\cite{Fl76} for a crystal with vacancies.
In the same limit $\Pi_{ik}$ coincides
with the normal expression for a momentum current density $\Pi_{ik} = \rL\veli v_{{\scs L}k} - \sigma_{ik} - \sigma^{\prime}_{ik}$
and in  the linear approximation in strains the expression (\ref{17}) is
reduced to the standard definition of  the energy current density in the viscoelastic medium
\[{\bf Q}  =  \bigl (\frac{\rho v_{\scs L}^2}{2} + E_0 \bigr )\vel - \vel\cdot({\bf \sigma} + \sigma^{\prime}) + {\bf q}  \, .\]
Having substituted the found expression for $\Pi_{ik}$ and ${\bf f}$  in (\ref{3})
and (\ref{6}) and restricting to the linear terms in lattice strains, we obtain the complete set of
the hydrodynamic equations  for a crystal with interstitials
\begin{eqnarray}
&&\frac{\partial\rho}{\partial t} + \mbox{div}\, {\bf j}  =  0\, ,   \nonumber \\
&&\rho\frac{\partial c}{\partial t} + ({\bf j}\cdot\nabla)\, c + \mbox{div}\, {\bf J}  =  0\, ,    \nonumber \\
&&\frac{\partial S}{\partial t} + \mbox{div}\,\Big ( \frac{S}{\rho}{\bf j} + \frac{{\bf q}}{T} - \frac{\nu}{\rho T}{\bf J}\Big ) = \frac{R}{T}\, ,   \label{23} \\
&& \rL\frac{\partial \vel}{\partial t} + \rL\velc\vel = - (1 - c)\nabla p      \nonumber \\
& &\phantom{\rL\frac{\partial \vel}{\partial t} + +  } + \nabla\cdot {\tilde{\bf \sigma}} + \nabla\cdot{\bf \sigma}^{\prime}+  c\rL\nabla \frac{w^2}{2} , \nonumber \\
& &\frac{\partial {\bf j}}{\partial t} +\vel\mbox{div}\, {\bf j} + ({\bf j}\cdot\nabla)\vel
+ {\bf j}_0\mbox{div}\,{\bf v}_p    \nonumber\\
& & \hspace{0.5cm}+ ({\bf v}_p\cdot\nabla){\bf j}_0 = - \nabla p + \nabla\cdot\tilde{\bf \sigma} + \nabla\cdot{\bf \sigma}^{\prime} \nonumber\\
& & \hspace{5cm}- \nabla\cdot{\bf \pi}\, ,    \nonumber
\end{eqnarray}
where $R$ and the dissipative currents are defined by expressions (\ref{18}) and
(\ref{20}).In addition, the system (\ref{23}) has to be still supplemented by the
conservation energy law (\ref{10}).

Now as a simplest example of the application of the obtained equations we shall consider
the propagation of a sound in the infinite cristal with the low interstitial concentration.
We shall assume that  effects of the viscosity and the heat conductivity
are absent so that the only dissipative process is related to the interstitial diffusion.
In addition, we shall use the isotropic approximation for both the elastic stress tensor
and the diffusion current, taking
\[\tsi_{ij} = 2\mu_0\tilde{u}_{ij}\, , \hspace{0.5cm}{\bf J} = - T\beta\nabla\Big(\frac{\nu}{\rho T}\Big)\, ,\]
where $\mu_0$ is the shear modulus and the constant $\beta$ is proportianal to
the diffusion coefficient. Further we shall confine ourselves to the case of the
small diffusion mobility of interstitials when $\beta \ll 1$. Assuming  that
a plane wave propagates along the $x$-direction and having eliminated the current
${\bf j}$, one writes the set of the linearized
equations (\ref{23}) in the form
\begin{eqnarray}
&&{\ddot{\rho}}^{\prime} = \frac{\partial^2 p^{\prime}}{\partial x^2} - \frac{4}{3}\mu_0\frac{\partial^3 u_x}{\partial x^3} \, , \nonumber \\
&&\rho{\dot c}^{\prime} = T\beta \frac{\partial^2 \onu ^{\, \prime}}{\partial x^2} \, ,\hspace{0.8cm}(\onu = \nu/\rho T)\, , \nonumber \\
&& \rho\dot s^{\prime} = - \frac{\nu}{\rho}\frac{\partial^2 \onu ^{\, \prime}}{\partial x^2} \, , \hspace{1cm}     ( s = S/{\rho})\, ,  \nonumber\\
&&\rL{\ddot u}_{x} = -(1 - c)\frac{\partial p^{\prime}}{\partial x} + \frac{4}{3}\mu_0\frac{\partial^2 u_x}{\partial x^2}\, ,\label{24}    \\
&& \ddot{u}_y = c_t^2\frac{\partial^2 u_y}{\partial x^2}\,, \hspace{0.9cm}    \ddot{u}_z = c_t^2\frac{\partial^2 u_z}{\partial x^2} \, .\nonumber
\end{eqnarray}
Here the prime denotes small deviations of a variable from its equilibrium value
which is without prime and $s = S/\rho$ is the entropy per unit mass.
It follows from the last two equations that in the assumed approximation the sound velocity of the
transverse waves $c_t^2 = \mu_0/\rL$  does not depend on the interstitial concentration
and coincides with its usual value in the isotropic elastic medium.

The equality (\ref{21}) shows that $p$, $c$, $T$, $w^2$ and $\tsi_{ij}$
can be taken as the independent variables. Taking into account the fact that the
scalar function can depend on the tensor $\tsi_{ij}$  by convolution $\tsi_{ij}\tsi_{ij}$
only,
in the linear approxomation, one can consider $\rho$, $\onu$, $s$ to be functions of $p$, $c$, and $T$.
In the following we shall take the space and time dependence of all variables to be
of the form $exp[i\omega(t + x/v)]$, where $v$ is the longitudinal sound velocity ,
and write down the set (\ref{24})  as
\begin{eqnarray}
&&v^3\frac{\partial\rho}{\partial T}T^{\prime} + \bigl ( v^3\frac{\partial \rho}{\partial p} - v\bigr )p^{\prime} + v^3\frac{\partial \rho}{\partial c}c^{\prime}  \nonumber \\
&&\phantom{v\frac{\partial\rho}{\partial T}T^{\prime} + \bigl ( v^3\frac{\partial \rho}{\partial p} - v\bigr )p  }+ \frac{4}{3}\mu_0i\omega u_x = 0\, ,   \nonumber\\
&&i\omega T\beta(\frac{\partial\onu}{\partial T}T^{\prime} + \frac{\partial\onu}{\partial p}p^{\prime}) + (i\omega T\beta\frac{\partial\onu}{\partial c} - v^2\rho)c^{\prime} \nonumber\\
&& \phantom{(\frac{\partial\onu}{\partial T}T^{\prime} + \frac{\partial\onu}{\partial p}p^{\prime}) + (i\omega T\beta\frac{\partial\onu}{\partial c} - v^2\rho) } =  0 \, ,  \nonumber \\
&& \frac{\partial s}{\partial T}T^{\prime} + \frac{\partial s}{\partial p}p^{\prime} + (\frac{\partial s}{\partial c} + \onu )c^{\prime}  = 0 \, ,\nonumber\\
&&  (1 -c)vp^{\prime} + i\omega (\rL v^2 - \frac{4}{3}\mu_0)u_x = 0\, .   \nonumber       \\
\label{25}
\end{eqnarray}
We shall find the sound velocity in the form $v = v_0 + \beta v_1$, where $\beta\rightarrow 0$.
Having put in  (\ref{25}) $\beta = 0$ one finds zero approximation $v_0$ as
\be
v_0^2 = \frac{4}{3}\frac{\mu_0}{\rL} + \Bigl ( \frac{\partial s}{\partial T}\bigr )_p \biggm /\frac{\partial (\rho , s)}{\partial (p, T)} + O(c) \, , \label{26}
\ee
where $O(c)$ denotes the small terms to be proportianal a concentration $c$. Using
the well known properties of a functional determinants, one has
\be
\frac{\partial (\rho , s)}{\partial (p, T)} =  \frac{\partial (\rho , s)}{\partial (p, s)} \frac{\partial (p , s)}{\partial (p, T)}
= \Big (\frac{\partial \rho}{\partial p}\Big )_s\Big (\frac{\partial s}{\partial T}\Big )_p. \label{27}
\ee
Since the adiabatic compression modulus per unit mass $K_{ad}$ is defined  as
\[\frac{1}{K_{ad}}= \frac{1}{\rho}\Big (\frac{\partial \rho}{\partial p}\Big )_s\, ,\]
we have, instead of (\ref{27})
\[\frac{\partial (\rho , s)}{\partial (p, T)} = \frac{\rho}{K_{ad}} \Big (\frac{\partial s}{\partial T}\Big )_p\, .\]

Substitution of the last equality in the Eq.(\ref{26}) yields
\be
v_0^2 = \frac{4}{3}\frac{\mu_0}{\rL} + \frac{K_{ad}}{\rL}\, ,\label{28}
\ee
since $\rho\rightarrow\rL$ at $c\rightarrow 0 $. The expression (\ref{28})
coincides with usual value of the longitudinal sound velocity in the isotropic elastic medium.

The compatibility condition of the set of Eqs.(\ref{25}) up to terms of the second
order in $\beta$ yields an equation for $v_1$. To avoid too combersome expressions we
use a number of simplifying assumptions. It is known that when introducing an impure particle
into a perfect crystal its volume changes in a macroscopic value. Therefore one can
suppose that at $c\rightarrow 0$ $\partial\rho/\partial c $  has to be considerably
more than all the other thermodynamic derivatives included into the set (\ref{25}).
In addition, we shall confine ourselves to the temperature region in which one
can ignore a heat expansion and neglect by terms containing ${\partial \rho}/{\partial T}$.
Then  keeping in an expression for $v_1$ only terms to be proportianal to
$\partial\rho/\partial c$, we obtain
\begin{eqnarray}
v_1 &=& i\omega\frac{T\delta}{2\rL v_0}\Big (\frac{\partial \rho}{\partial c}\Big )_{pT}\, , \nonumber
\end{eqnarray}
where
\begin{eqnarray}
\delta & = & \frac{\partial (\onu , s)}{\partial (p, T)} \Big(\frac{ \partial s}{\partial T}\Big)_{cp}^{-1} \nonumber\\
&& \times \frac{2K_{ad}}{ (2\rL v_0^2 + 3 K_{ad})({\partial \rho}/{\partial p})_{cT} - 3\rL  }. \nonumber
\end{eqnarray}
Allowing for $v = v_0 + \beta v_1$, one has for the absorption coefficient
\be
\gamma_l = \frac{\omega^2 T \beta\delta}{2\rL v_0^3}\Big(\frac{\partial \rho}{\partial c}\Big)_{pT}  \label{29}
\ee
As it seen from Eq.(\ref{29}) $\gamma_l$ to be proportianal to $\omega^2$ and the diffusion
coefficient.

In conclusion, we have obtained the complete system of the hydrodynamic equations for a
crystal with interstitials. These equations allow to describe the joint
dynamics of the lattice and the impurity system and take into
account the dissipative efects of the
viscosity, the heat conductivity and the
diffusion of interstitials. On the basis of found
equations we have considered the problem
of the propagation of plane waves in the crystal
with small interstitial concentration and calculated the sound velocities
and absorption coefficient provided that
the viscosity and the heat conductivity are
absent and the diffusion mobility of interstitials is small.

%
%

%
%

\end{document}